\documentclass[pdflatex,sn-mathphys-num]{sn-jnl}

\usepackage{graphicx}%
\usepackage{multirow}%
\usepackage{amsmath,amssymb,amsfonts}%
\usepackage{amsthm}%
\usepackage{mathrsfs}%
\usepackage[title]{appendix}%
\usepackage{xcolor}%
\usepackage{textcomp}%
\usepackage{manyfoot}%
\usepackage{booktabs}%
\usepackage{algorithm}%
\usepackage{algorithmicx}%
\usepackage{algpseudocode}%
\usepackage{listings}%
\usepackage{caption}
\usepackage{tabularx}
\usepackage{float}
\usepackage[numbers,sort&compress]{natbib}
\usepackage{pdflscape}
\usepackage[strict]{changepage}
\newcolumntype{C}{>{\centering\arraybackslash}X}

\theoremstyle{thmstyleone}%
\theoremstyle{thmstyletwo}%

\theoremstyle{thmstylethree}%

\begin{document}

\title[Article Title]{Weather-Related Crash Risk Forecasting: A Deep Learning Approach for Heterogenous Spatiotemporal Data}


\author[1]{\fnm{Abimbola} \sur{Ogungbire, Ph.D.}}\email{abim.ogungbire@greshamsmith.com}

\author*[2]{\fnm{Srinivas} \sur{Pulugurtha, Ph.D., PE}}\email{sp@olemiss.edu}

\affil[1]{\orgdiv{Traffic Engineering}, \orgname{Gresham Smith}, \orgaddress{\street{Alpharetta}, \postcode{30009}, \state{GA} - \country{USA}}}

\affil*[2]{\orgdiv{Civil Engineering}, \orgname{University of Mississippi}, \orgaddress{\street{Oxford}, \postcode{38677}, \state{MS} - \country{USA}}}


\abstract{This study introduces a deep learning-based framework for forecasting weather-related traffic crash risk using heterogeneous spatiotemporal data. Given the complex, non-linear relationship between crash occurrence and factors such as road characteristics, and traffic conditions, we propose an ensemble of Convolutional Long Short-Term Memory (ConvLSTM) models trained over overlapping spatial grids. This approach captures both spatial dependencies and temporal dynamics while addressing spatial heterogeneity in crash patterns. North Carolina was selected as the study area due to its diverse weather conditions, with historical crash, weather, and traffic data aggregated at 5-mi by 5-mi grid resolution. The framework was evaluated using Mean Squared Error (MSE), Root Mean Squared Error (RMSE), and spatial cross-K analysis. Results show that the ensembled ConvLSTM significantly outperforms baseline models, including linear regression, ARIMA, and standard ConvLSTM, particularly in high-risk zones.  The ensemble approach effectively combines the strengths of multiple ConvLSTM models, resulting in lower MSE and RMSE values across all regions, particularly when data from different crash risk zones are aggregated. Notably, the model performs exceptionally well in volatile high-risk areas (Cluster 1), achieving the lowest MSE and RMSE, while in stable low-risk areas (Cluster 2), it still improves upon simpler models but with slightly higher errors due to challenges in capturing subtle variations.}

\keywords{Crash Forecasting, Spatiotemporal, LSTM, Weather-related Crash, Deep learning, Long-short term memory, Ensembled-ConvLSTM}



\maketitle

\section{Introduction}\label{sec1}

Fast urbanization in most US cities has introduced both safety and sustainability challenges. These challenges are even worse in states with epileptic weather conditions, such as North Carolina. The federal highway administration (FHWA) reported that 21 percent of all crashes are weather-related \cite{bib1}. Between 2007 and 2016, nearly 5,400 people were killed in weather-related crashes, making it one of the main contributors to traffic crashes \cite{bib1}.Thus, Intelligent Transportation Systems (ITS) has become an active research area given its potential to reduce crashes in poor weather conditions \cite{bib2}. As an essential step towards improving the ITS, weather-related crash forecasting aims at projecting into the future crash status at specific location within a traffic system.

Spatial analysis of crashe data is becoming more increasingly popular. In the last decades, researchers have made considerable effort in analyzing crash data at various spatial levels \cite{bib3, bib4, bib5, bib6}. In our most recent work \cite{bib5}, \cite{bib46}, we set a precedence for space-time cube theory in crash risk analysis. We detailed the preference of state agencies to examine crash data at spatially aggregated levels, including traffic analysis zones (TAZ) \cite{bib7} and grid-level spatial units \cite{bib5}, \cite{bib8}, to easily facilitate effective resource allocation \cite{bib9}. In this study, we build upon the groundwork laid in \cite{bib5} to forecast potential future risk of weather-related crash events at grid level.

Recent surge in the availability of large datasets related to human activities in urban settings has spurred a significant increase in research on traffic incidents \cite{bib10, bib11, bib12}. This wealth of data offers an unprecedented opportunity to learn from historical events to better understand and predict future traffic incidents. Recent research efforts have explored integrating big data into spatially aggregated crash models \cite{bib10}, \cite{bib13}. A study by Bao et al.\cite{bib10} investigated the use of big data derived from traffic sensors and social media to improve spatially aggregated crash models. They developed a methodology that combines traditional traffic data with real-time social media analytics to predict crash hotspots in urban areas. 

Early studies have been formulated to view traffic crash forecasting as either a classification or a regression problem. For example, some work aims to predict the likelihood of a crash occurrence at specific location \cite{bib14, bib15} or time periods \cite{bib15}. Looking at crash forecasting from this lens allow researchers to identify significant predictors of crashes and quantify their impacts. Conversely some studies are focused on estimating the intensity of crashes at specific location during each time window \cite{bib13, bib17, bib18, bib19}. Over time, the field has seen the integration of more sophisticated approaches, including time-series analysis \cite{bib20}, \cite{bib21} for understanding temporal patterns and machine learning models \cite{bib22}, \cite{bib23} for capturing complex, non-linear relationships between variables. Geographic information systems (GIS) have also been applied to spatially analyze crash data and identify high-risk areas \cite{bib24, bib48}. These methods have evolved from simple, deterministic models to dynamic, probabilistic models that better account for the uncertainties inherent in predicting human behavior and environmental interactions.

Deep learning approach, which is gaining widespread popularity in computer vision \cite{bib25}, natural language processing \cite{bib26}, artificial intelligence, and pattern recognition \cite{bib25}, \cite{bib27}, is now being to be applied in traffic safety research. This includes applications such as traffic conflict prediction \cite{bib27}, \cite{bib28}, near miss identification at intersections \cite{bib29}, estimating unsafe driving speed, and red-light violation at signalized intersection \cite{bib30}. Deep learning distinguishes itself from traditional statistical models and other learning architectures by its ability to model complex non-linear relationships through distributed and hierarchical feature representation \cite{bib31}, demonstrating superior performance in predicting short-term traffic flow and speed. 

In this paper, we introduce an ensembled convolutional long short-term Memory (ensembled-ConvLSTM), a deep learning method designed to predict traffic crashes using big and heterogenous spatio-temporal data. We divide the study area into grid cells and collect detailed information on urban and environmental features such as traffic volume, road conditions, traffic management data, and weather conditions for each cell. Using this data, we develop a model to forecast the number of crashes that will occur in each grid cell in future time periods. Our approach utilizes a ConvLSTM neural network, which helps us to better understand and predict patterns and differences in the data over time and space. In addition, we create an ensemble framework in which separate models are trained for different risk regions and their results are combined to produce the final prediction. To our knowledge, this is the first study to address spatial variability in traffic crash prediction using deep neural networks and the first to apply ConvLSTM for this purpose. Our results show that our model significantly outperforms traditional methods, achieving higher accuracy in predicting traffic crashes.

\section{Related Works}\label{sec1}

We begin our discussion in this section by describing the role of poor weather conditions in road crashes. Next, we describe the consensus on the use of traditional models for traffic crash forecasting and the need for a classical approach. Finally, we give an up-to-date analysis of deep learning techniques for traffic crash prediction.

\subsection{Weather-Related Crashes}\label{subsec2}

Crashes are a result of multiple factors that can be categorized into behavioral, technological, and environmental influences \cite{bib23}. The absence of any of these elements has the potential to prevent crashes. While weather is not the primary cause of road crashes \cite{bib1}, \cite{bib23}, its significance cannot be overlooked. Studies, including those by Vickery \cite{bib32} and Downs \cite{bib33}, indicate that most people do not consider poor weather as a deterrent to driving unless conditions severely impede travel. Bergel-Hayat et al. \cite{bib34} among others, have established a correlation between weather conditions and road transport, detailing how adverse weather can lead to inconvenience or even compel travelers to cancel their travel plan \cite{bib34}.

Vehicles, unlike other modes of transport, are generally not designed to operate under extreme weather conditions. This contrasts with aircraft, which are equipped to handle severe atmospheric conditions. The impact of bad weather on road safety is complex and cannot be reduced to simple cause and effect. Study by Jackson \& Sharif \cite{bib35} shows that rain increases crash rates, a situation exacerbated by more people choosing to drive under wet conditions. However, the introduction of technologies such as anti-lock brakes and traction control has changed the dynamics of driving in poor weather, potentially leading to riskier driving behaviors as drivers gain confidence from these features \cite{bib36}. 

Interestingly, weather-related challenges do not always lead to severe crashes. In some situations, like snow, drivers tend to be more cautious, reducing their speed and thus mitigating risk. Decisions to cancel or postpone travel plans can also decrease the likelihood of crashes during unfavorable weather conditions \cite{bib37}.

\subsection{Crash Forecasting Using Traditional Models}\label{subsec2}

The evolution of traffic crash forecasting using traditional models shows a shift from linear statistical models to more dynamic and complex computational models \cite{bib14, bib15, bib17, bib18, bib19, bib20, bib21, bib22, bib23}. The integration of different forecasting techniques, such as combining grey models with Markov chains or enhancing ARIMA with neural network analysis for the non-linear components, exemplifies the interdisciplinary approach towards a more accurate and robust prediction of traffic incidents.

Time-series forecasting methods have been foundational in predicting traffic crashes, utilizing historical data to estimate future outcomes \cite{bib20}. These methods consider the sequence of data points collected over time, analyzing patterns such as long-term trends, seasonality, and irregular factors. Key approaches within time-series forecasting include the exponential smoothing method \cite{bib20}, \cite{bib38}, which emphasizes the diminishing significance of older data, and ARIMA (Autoregressive Integrated Moving Average), a model that integrates differencing of observations (to remove non-stationarity) with autoregression and moving averages \cite{bib20}, \cite{bib38}. These methods are based on the premise that past patterns in traffic crash data can offer insights into future occurrences, with techniques like exponential smoothing and ARIMA being particularly noted for their ability to model and predict traffic crashes \cite{bib20}, \cite{bib38}, \cite{bib39}.

Exponential smoothing models, including the simple exponential smoothing and its extensions to account for trend and seasonality \cite{bib38}, provide a framework for smoothing out time series data to identify underlying trends. The sophistication of these models lies in their statistical rationale, which accommodates various forms of trends and seasonality through state-space models \cite{bib39}. ARIMA and its precursor, ARMA (Autoregressive Moving Average), further advance the field by addressing the stochastic properties of time series and facilitating model selection based on the stationary characteristics of the data. These models have been applied to correct error terms in traffic crash forecasting, combining with other methodologies like regression models to enhance the reliability of predictions.

Markov chain models introduce a probabilistic approach to forecasting, emphasizing the transition probabilities between discrete states over time \cite{bib40}. This method suits scenarios with significant random fluctuations but without clear trends, offering insights into the stochastic nature of traffic crashes. 

\subsection{Deep Learning Models for Weather-Related Crash Predictions}\label{subsec2}

Deep learning, a subset of machine learning characterized by its use of neural networks with multiple layers, has emerged as a powerful tool in the realm of traffic crash forecasting \cite{bib25, bib26, bib27, bib28, bib29}. The application of deep learning in traffic crash forecasting in literature has been restricted to specific models such as convolutional neural networks (CNNs), recurrent neural networks (RNNs), and their variations such as long short-term memory (LSTM) networks and gated recurrent units (GRUs). These architectures are adept at handling the spatial and temporal data inherent in traffic systems, allowing for the modeling of complex patterns and relationships that traditional models might overlook.

CNNs have proved effective in processing spatial data, making them suitable for analyzing crash data aggregated by geographical units, such as grid maps \cite{bib25}. By capturing spatial dependencies through their convolutional filters, CNNs can identify patterns related to traffic flow, road infrastructure and other spatial factors contributing to crash risks \cite{bib29}. On the other hand, RNNs and LSTMs, are designed to handle sequential data, allowing for temporal dynamics of traffic crash occurrences. These models can learn from historical crash data, recognizing patterns over time, such as the cyclic nature of traffic volume and its correlation with crash incidents \cite{bib31}. In addition, LSTMs that can remember long-term dependencies, are particularly effective in overcoming the vanishing gradient problem common in traditional RNNs.

\begin{landscape}
\pagebreak
\thispagestyle{empty}
\vspace*{\fill} 

\begin{adjustwidth}{-1in}{-1in} 
\begin{center}
    \captionof{table}{Comparison of Crash Forecasting Techniques} \label{tab2}
    \addvspace{10pt}
    
    \small 
    \begin{tabularx}{1.0\linewidth}{@{} l >{\raggedright\arraybackslash}p{110pt} l >{\raggedright\arraybackslash}X >{\raggedright\arraybackslash}X @{}}
    \toprule \toprule
    \textbf{Authors} & \textbf{Methodologies} & \textbf{Range} & \textbf{Features used} & \textbf{Major improvement}\\
    \midrule
    Duddu \& Pulugurtha \cite{bib14} & Linear regression + BPNN & Short term & On network characteristics, and land use characteristics & Link-level crash frequency model developed. \\ \addlinespace
    Loo et al. \cite{bib25} & RF + XGBoost + Naïve bayes NB & Short term & Pedestrian exposure, jaywalking &Both XGBoost and RF models generated similar results on feature importance for three sets of models. Also, there are non-linear relationships of many risk factors with bus crashes \\ \addlinespace
    Formosa et al. \cite{bib28} & R-CNN & Short term & Speed, vehicle sensor data (yaw rate, velocity, longitudinal displacement, etc.), headway, occupancy & Predict traffic conflict by mining heterodox data. \\ \addlinespace
    Rabbani et al. \cite{bib38} & SARIMA + ES & Medium term & Historical crash frequency & Exponential smoothing model has a better fit on crash data over SARIMA judging from the MAE, RMSE, MAPE, and BIC score \\ \addlinespace
    Cai \& Di \cite{bib41} & ARIMA + boosting & Short term & Traffic flow, weather, speed & Integrating time series with a count data model can capture traffic crash features and account for the temporal autocorrelation \\ \addlinespace
    Ivan \cite{bib43} & Bayesian framework & Short term & -- & Traffic volume use in crash rate analysis. \\ \addlinespace
    Ladron de Guevara et al. \cite{bib44} & Negative binomial & Short term & Population, TAZ, schools & Model for equitable planning/incentive programs. \\ \addlinespace
    Huang et al. \cite{bib45} & Deep dynamic fusion & Short/Long & Time, road condition, blocked drive & Improved DNN modeling for dynamic conditions. \\ 
    \bottomrule \bottomrule
    \end{tabularx}
\end{center}
\end{adjustwidth}

\vspace*{\fill}
\end{landscape}

\subsection{Literature Gaps and Future Directions}\label{subsec2}

Despite the promise shown by deep learning in traffic crash forecasting, several challenges remain. These include the need for large and diverse datasets to train the models effectively, the computational complexity and resource requirements of deep learning algorithms, and the ‘black box’ nature of these models that often makes it difficult to interpret their predictions. Addressing these challenges requires ongoing research and development, including efforts to improve model transparency and interpretability, enhance computational efficiency, and ensure the ethical use of traffic data.

Furthermore, the integration of deep learning models with emerging technologies, such as the Internet of Things (IoT) and edge computing, presents exciting opportunities for the future of traffic crash forecasting. By leveraging real-time data collection and processing at the edge of the network, it is possible to develop more dynamic and responsive forecasting models that can adapt to changing traffic conditions and contribute to the development of safer and more efficient transportation systems.

\section{Methodology}\label{sec1}

This section presents the data used in the study, introduces the formulation of our problem, and present our feature extraction technique.

\begin{figure}[h]
    \centering
    \includegraphics[scale = 0.4]{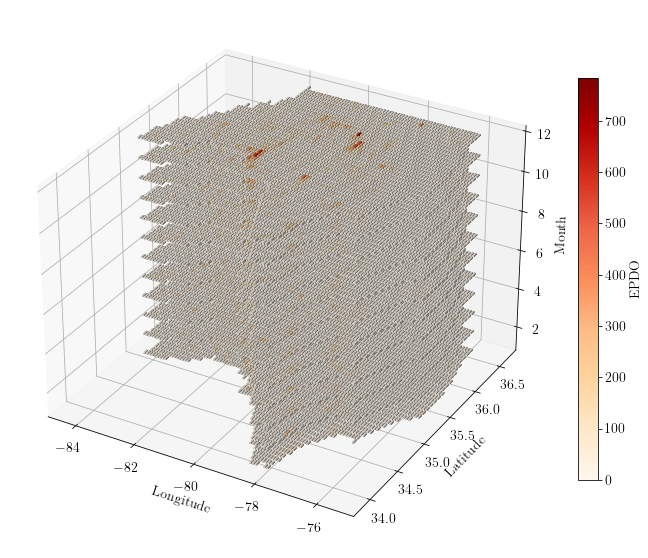}
    \caption{Spatiotemporal cube constructed in the study area}
    \label{fig:cube}
\end{figure}

\subsection{Data Sources}\label{subsec2}

For our analysis, we selected North Carolina (NC), USA, as the study area. NC is characterized by its wide range of weather phenomena, including but not limited to snowfalls, rainfall, and wind. This diversity in weather conditions makes NC an exemplary state for examining the effects of weather, particularly precipitation, on road traffic incidents, a point made in a study by \cite{bib42} in their research. The primary dataset for our investigation was sourced from the highway safety information systems HSIS \cite{bib43, bib47}, encompassing vehicle crash records spanning 2015 through 2018. Pre-COVID data were used because pandemic-era travel behavior introduced structural breaks in traffic demand, exposure, and crash mechanisms. Including such disrupted periods would violate the stationarity assumptions underlying spatiotemporal forecasting models and bias parameter learning, reducing generalizability to normal roadway conditions. Building on the work of \cite{bib5, bib46}, this study extends the application of the space-time cube framework to forecast the potential future risk of weather-related traffic crashes at the grid level. The spatial distribution of crash locations across NC is presented in Figure \ref{fig:cube}. Weather features were excluded from the model to prevent information leakage. For operational use of the model, future weather for the prediction week will not be available in real time, hence, the model was trained using only crashes recorded under inclement weather conditions, allowing prediction of weather-related crash risk without using future weather data.

\subsection{Problem Definition}\label{subsec2}

\begin{figure}[h]
    \centering
    \includegraphics[scale = 0.8]{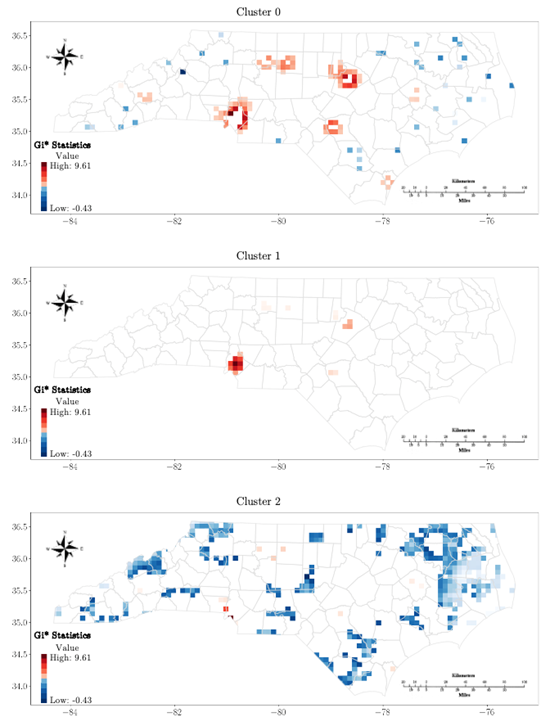}
    \caption{DTW-G* Clusters}
    \label{fig:cluster}
\end{figure}

Our objective is to construct a predictive framework that estimates the total number of traffic crashes within specified units of a spatial grid $S$ over distinct time intervals. This grid, denoted as $S=\{s_1,s_2,\dots,s_n\}$, comprises $n$ subdivisions, each representing an area of $d \times d$ square miles. For illustrative purposes, consider $d=5\text{mi}$, whereby the entire geographical expanse of North Carolina could be dissected into a grid formation of $2,045$ units. Time is segmented into discrete intervals, referred to as slots, with a week ($7$ days) being the standard length for this analysis, albeit the methodology supports adjustments in both spatial ($d$) and temporal ($t$) dimensions. Figure \ref{fig:cluster} shows the EPDO trends of the training dataset. We formulate the problem as follow:

\vspace{0.5em}
\textbf{Provided Inputs:}
\begin{itemize}
    \item A spatiotemporal framework encapsulated by a matrix $S \times T$, where $S$ symbolizes the spatial grid with its $n$ divisions, and $T$ embodies the sequence of $t$ time intervals during the study period.
    \item A matrix $A$ of dimensions $n \times t$, with element $A_{ij}$ representing the crash EPDOs in spatial unit $s_i$ at time $t_j$.
    \item A series of $m$ matrices $\{M_1, M_2, \dots, M_m\}$, with each matrix $M_k$ of dimensions $n \times t$, capturing distinct attributes pertinent to each grid unit $s_i$ over the time slots $t_j$.
    \item Training data set $D_{\text{train}}$ composed of pairs from $A$ and the feature matrices for time slots within $T_{\text{train}}$, and a testing data set $D_{\text{test}}$ containing pairs for time slots in $T_{\text{test}}$.
\end{itemize}

\vspace{0.5em}
\textbf{Objective:}
\begin{itemize}
    \item Formulate a model that can accurately forecast the crash count matrix $A$ for all time intervals $t_j \in T_{\text{test}}$, aiming to minimize the discrepancy between the predicted and actual crash counts.
\end{itemize}

\vspace{0.5em}
\textbf{Constraints:}
\begin{itemize}
    \item The correlation between crash counts and features ($M_k$) varies across different spatial units.
    \item Crashes are presumed to occur exclusively within the confines of the road network.
    \item For any forthcoming timeslot $t_i$, the corresponding feature matrices $M_{k,t_i}$ are not accessible for use in predicting $A_{i,t_i}$, signifying $t_i \in T_{\text{test}}$.
\end{itemize}

\subsection{Feature Extraction}\label{subsec2}

To prepare the features for our model, the dataset was aligned with each grid $s_i$ and week $t_i$ combination, aggregating the data to compile a list of features. For the dependent variable, we calculated the EPDO for each grid $s_i$ on each week $t_i$, from January 2015 to December 2018. The process for extracting the independent features is detailed in the sections below.

We mapped the road network onto grid cells, overlaying it with a mask layer to delineate the study area. It's important to note that traffic crashes are restricted to the road network, despite the grid-based partitioning of the entire area. We assessed the risk level in each grid based on two factors: crash frequency and crash severity. For more accurate predictions, we normalized the EPDO score for each grid by the total road length within that grid, assigning null values to grids without roads. Given the stability of the road network over time, this feature is considered time-invariant.

In our analysis, we augmented the network mask layer by calculating and storing two additional measures for each grid cell: the average road length and the average speed limit. Further, we incorporated features associated with road properties, which include the proportion of different traffic control types, average number of lanes, proportion of different route types, road, and Annual Average Daily Traffic (AADT) with each grid cells $s_i$. These features are considered time-invariant.

\subsection{Experimental Settings}\label{subsec2}

This study aims to forecast weather-related crashes in North Carolina for the year 2018 using a deep learning model trained on data from 2015 to 2017. The dataset is aggregated weekly, resulting in a sequence of 209 time frames: 157 for training, 52 for testing, and 10\% of the training frames set aside for validation to tune hyperparameter. Each time frame consists of a spatial tensor representing 2,153 5-mile grid cells across North Carolina.

For each week in 2018, the proposed ConvLSTM model predicts crash outcomes in each grid cell based on historical crash, road, and traffic data. Our research is guided by three research questions:

\begin{enumerate}[label=\roman*]
    \item Does the proposed framework outperform conventional models and the standard ConvLSTM in predictive accuracy?
    
    \item What variations in performance does our proposed model exhibit across different crash risk zones, such as areas of high and low risk?
    
    \item Do the model's predictions align spatially with actual crash locations, thereby confirming their logical validity?
\end{enumerate}

To assess the models' precision, we employ mean squared error (MSE) and root-mean-square error (RMSE) as our primary metrics. Furthermore, the Cross-K function serves to evaluate the spatial correlation between the predicted outcomes and the actual data.

\subsection{Spatiotemporal Ensembled-ConvLSTM}\label{subsec2}

\begin{figure}[h]
    \centering
    \includegraphics[scale = 0.3]{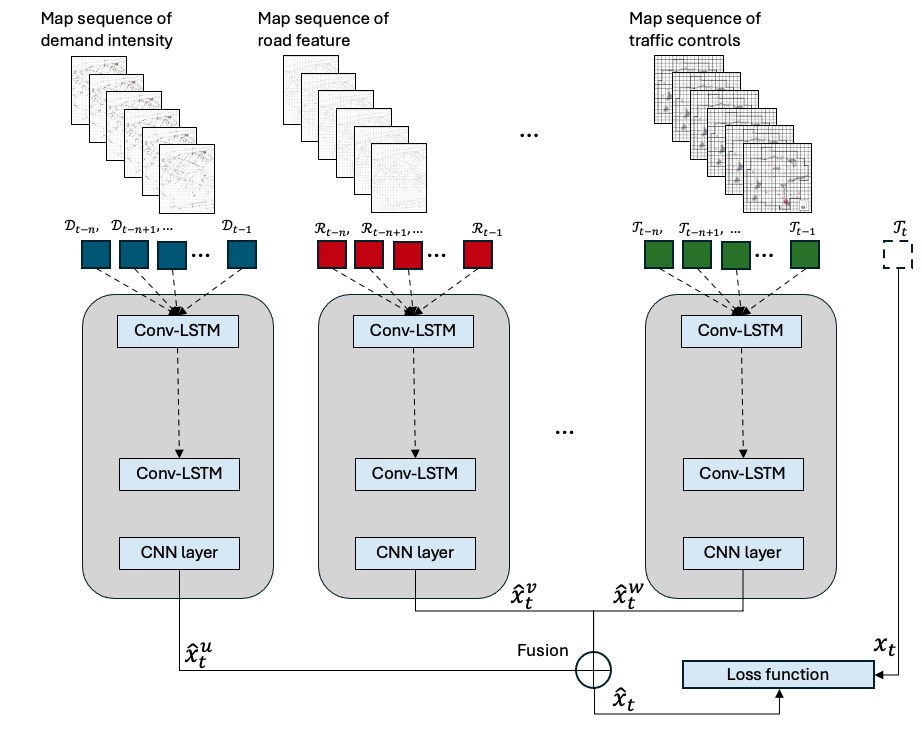}
    \caption{Single ConvLSTM Architecture}
    \label{fig:arch}
\end{figure}

In this section, we introduce our approach to forecasting weather-related crashes. Initially, we describe a single Convolutional LSTM (ConvLSTM) framework for our study. Subsequently, we elaborate on the construction of our Spatiotemporal Convolutional LSTM (Spatial-ConvLSTM) model, which incorporates multiple ConvLSTMs. The ConvLSTM model, an extension of the traditional LSTM, was initially developed by Shi et al (2015) for precipitation nowcasting. It is particularly suited for handling data where both spatial and temporal dimensions are crucial. Each input to the ConvLSTM network is treated as a 3D spatiotemporal tensor. The typical LSTM node is modified in the ConvLSTM to include convolution operations within its structure, as illustrated in a single ConvLSTM shown in Figure \ref{fig:arch}. Specifically, the input-to-state and state-to-state transitions in a ConvLSTM cell involve convolutional operations which output 3-D tensors. These modifications are governed by algorithm \ref{alg:convlstm}.

To tackle the challenge of spatial heterogeneity in weather-related crash prediction, the approach involves constructing distinct LSTM models for various clusters within the study area. These clusters are determined based on the spatial heterogeneity of the data, which can reflect varying risk levels, such as high-risk or low-risk zones as shown in Figure 4. The ensemble method is then applied to integrate the results from multiple models, thereby mitigating the effects of data heterogeneity.

\begin{algorithm}[H]
\caption{Spatial ensembled-ConvLSTM}
\label{alg:convlstm}
\begin{algorithmic}[1] 
\State Initialize model parameters for each window
\For{training epoch in $num\_epochs$}
    \For{each window in $study\_area$}
        \State Extract spatiotemporal data $x$ within window
        \For{each timestep $t$ in $x$}
            \State Compute input gate:
            \State $i_t = \sigma(w_{xi} * X_t + w_{hi} * h_{t-1} + b_i)$
            \State Compute forget gate:
            \State $f_t = \sigma(w_{xf} * X_t + w_{hf} * h_{t-1} + b_f)$
            \State Compute output gate:
            \State $o_t = \sigma(w_{xo} * X_t + w_{ho} * h_{t-1} + b_o)$
            \State Update cell state:
            \State $C_t = f_t \cdot C_{t-1} + i_t \cdot \tanh (w_{xc} * X_t + w_{hc} * h_{t-1} + b_c)$
            \State Hidden state output:
            \State $h_t = o_t \cdot \tanh (C_t)$
            \State Store last hidden state $h_t$
        \EndFor
        \State Aggregate and store output from all frames
    \EndFor
    \State Use ensembled method to combine predictions from all windows
    \State Evaluate model performance
    \State Adjust parameters based on gradients and learning rate
\EndFor
\end{algorithmic}
\end{algorithm}

A moving window technique was employed with each window measuring 10×10 grid cells. This is to help capture heterogeneous local meteorology pattern The study area is segmented by shifting this window from the upper-left corner at coordinates (0,0) to the bottom-right at coordinates (128, 64). The windows are moved across the grid in steps of 16 units both horizontally and vertically, ensuring comprehensive coverage and overlap across the study area. This strategy enabled the capture of localized spatial features significant for accurate crash prediction.

\begin{figure}[h]
    \centering
    \includegraphics[scale = 0.85]{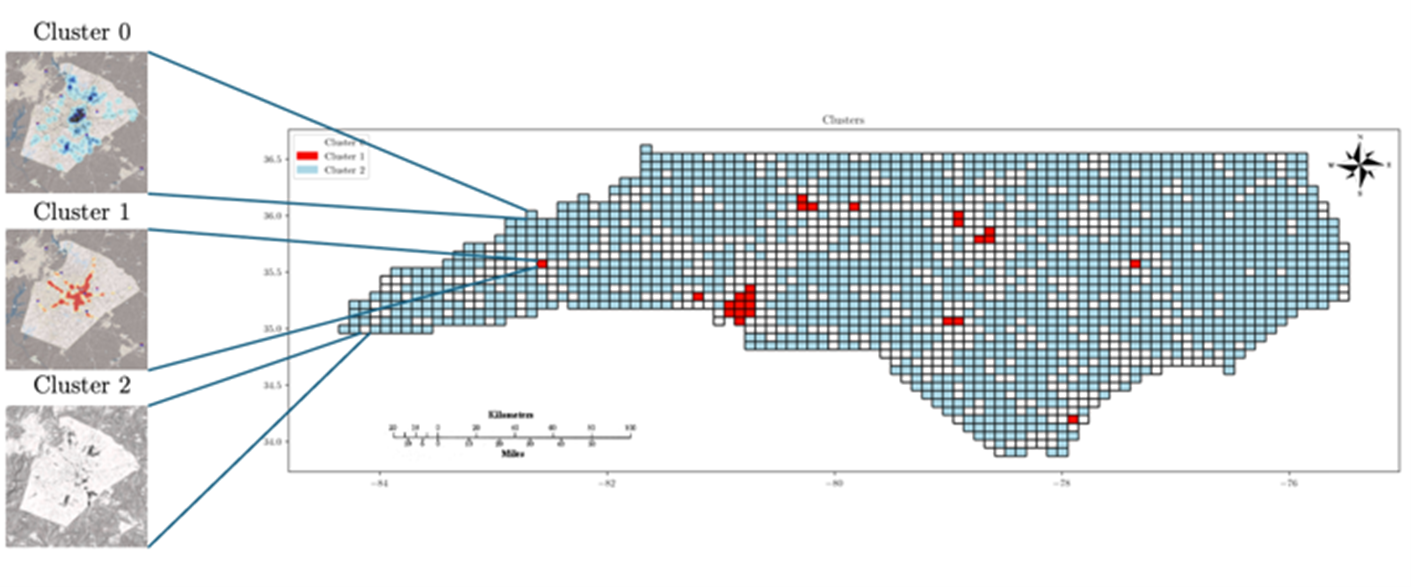}
    \caption{Distribution of Frames by Clusters}
    \label{fig:frame}
\end{figure}

For each windowed region, a dedicated ConvLSTM model was trained using the local training dataset corresponding to that window. The model then performed predictions on the testing dataset for that specific region. By training individual models on localized data, the unique spatial-temporal characteristics of each region were captured, which might be crucial due to varying meteorological and traffic conditions. The final prediction for a specific grid location $s_i$ on weeks $t_j$ is computed using an ensemble method, which aggregated predictions from all models covering $s_i$. This aggregation is performed as a weighted average of the predictions for $s_i$ at $t_j$  from all significant models as shown in (Equation \ref{eq1}). 

\begin{equation}
\hat{C}(s,t) = \frac{1}{\sum_{k=1}^{N} w_k} \sum_{i=1}^{N} w_i \hat{C}_i(s,t) \times I(s \in W_i). \label{eq1}
\end{equation}

\section{Results}\label{sec1}

\begin{figure}[h]
    \centering
    \includegraphics[scale = 0.7]{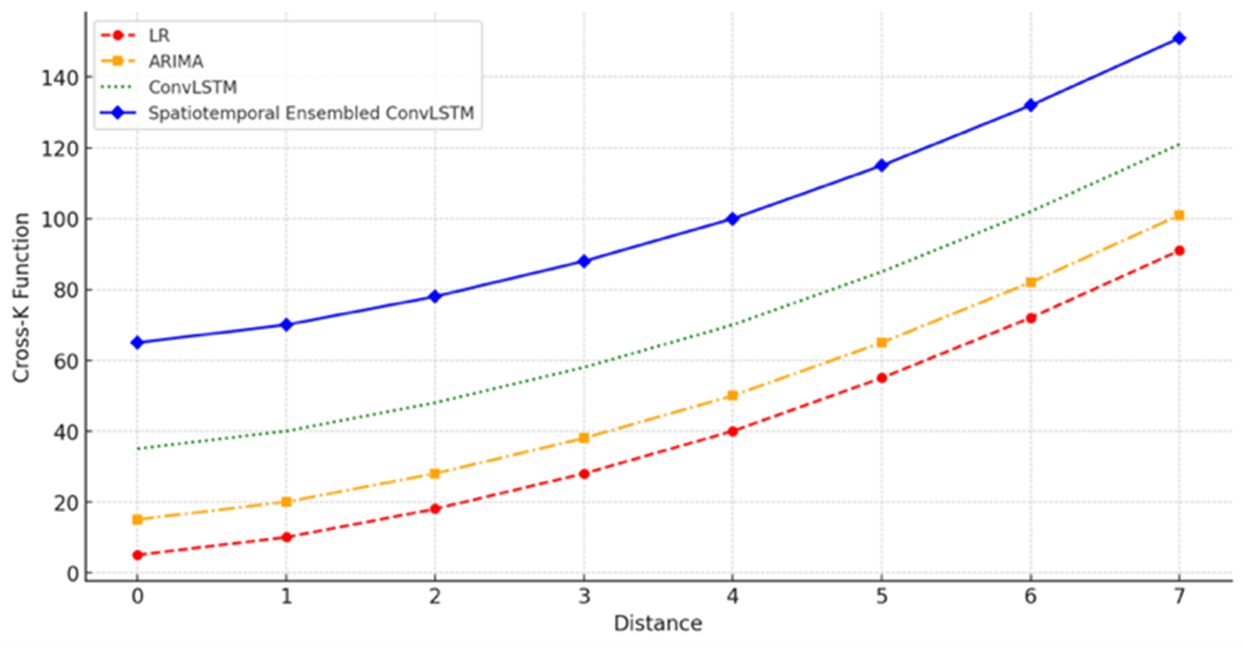}
    \caption{Cross-K function between predicted and actual weather-related crash risks}
    \label{fig:res}
\end{figure}

Figure \ref{fig:res} presents a comparison of four predictive models: LR, ARIMA, ConvLSTM, and Spatiotemporal Ensembled-ConvLSTM, using the cross-K function to measure their accuracy in forecasting weather-related crashes. The cross-K function values, plotted against ‘distance,’ serve to evaluate how closely each model’s predictions align with actual events.

LR is observed to have the lowest performance among the four predictive models, with its cross-K function values consistently rising but lower than other models throughout the range. This shows that while LR can predict trends, its simplicity restricts its effectiveness in capturing complex patterns in weather-related crash data. ARIMA has better performance than LR, yet still falls short compared to the neural network-based models. Its ability to incorporate past values and forecast errors into future predictions does provide an edge over LR, yet it lacks the capability to effectively handle spatial or multidimensional temporal dependencies, which are crucial in the context of weather-related events.

ConvLSTM substantially outperformed LR and ARIMA, underscoring the advantages of integrating convolutional layers into LSTM networks. This architecture enables the model to capture spatial features and temporal sequences simultaneously, which is particularly beneficial for modeling scenarios like weather patterns where both spatial and temporal dynamics are significant. Spatiotemporal Ensembled-ConvLSTM has the highest cross-K function values across all distances. This model combines multiple ConvLSTM models to leverage diverse spatial and temporal features more robustly, reducing the risk of overfitting to patterns and improving generalization across various scenarios.

The differences in performance can be attributed to several factors, for example, the model complexity and architecture, i.e., more complex models (ConvLSTM and spatiotemporal ensembled-ConvLSTM) are designed to handle the intricacies of spatial and temporal data simultaneously, which is crucial for accurately modeling phenomena like weather-related crashes that exhibit both spatial and temporal variability. In addition, the data handling capabilities of each model plays a role. The ability of ConvLSTM to process data in both time and space allows for a more nuanced understanding of how weather conditions across different regions influence crash rates over time. Furthermore, the superior performance of the spatiotemporal ensembled-ConvLSTM suggests that ensemble techniques, which combine predictions from multiple models to improve accuracy are particularly effective in dealing with complex, noisy datasets like those involving traffic.

\begin{center}
    \captionof{table}{Model Performance Evaluation}
    \label{tab:performance}
    \setlength{\tabcolsep}{0pt} 
    \begin{tabular*}{\textwidth}{@{\extracolsep{\fill}} l cccccccc @{}}
        \toprule \toprule
        \multirow{2}{*}{Model} & \multicolumn{2}{c}{Cluster 0} & \multicolumn{2}{c}{Cluster 1} & \multicolumn{2}{c}{Cluster 2} & \multicolumn{2}{c}{All regions} \\ 
        \cmidrule{2-3} \cmidrule{4-5} \cmidrule{6-7} \cmidrule{8-9}
         & RMSE & MSE & RMSE & MSE & RMSE & MSE & RMSE & MSE \\ \midrule
        LR & 0.321 & 0.103 & 0.146 & 0.021 & 0.184 & 0.034 & 0.852 & 0.7259 \\
        ARIMA & 0.288 & 0.082 & 0.091 & 0.008 & 0.151 & 0.023 & 0.543 & 0.2948 \\
        ConvLSTM & 0.253 & 0.064 & 0.073 & 0.005 & 0.084 & 0.007 & 0.331 & 0.1096 \\
        Ensembled-ConvLSTM & -- & -- & -- & -- & -- & -- & 0.024 & 0.0006 \\
        \bottomrule \bottomrule
    \end{tabular*}
\end{center}

Table \ref{tab:performance}. provides a comparative analysis of the mean squared error (MSE) and root mean squared error (RMSE) across different clusters for the four predictive models: LR, ARIMA, ConvLSTM, and spatiotemporal ensembled-ConvLSTM. Each cluster represents different characteristics of EPDO, which measures the severity and frequency of crashes. 

Cluster 0 represents gradually increasing low EPDOs. Here, LR, ARIMA, and ConvLSTM show progressively lower MSE and RMSE, indicating increasing accuracy with more sophisticated models. The improvement from LR to ARIMA and further to ConvLSTM suggests that the gradual increase in EPDO severity over time in this cluster is better modeled by algorithms that can handle time series data with trend and seasonality. However, cluster 1 represents volatile high-EPDOs. This cluster displays the lowest MSE and RMSE across all models, which may seem counterintuitive given its volatility. However, this can indicate that the models, particularly ConvLSTM, are effectively capturing the rapid fluctuations in EPDOs. The lower error metrics suggest that sophisticated models like ConvLSTM are particularly adept at managing the high variability within this cluster. Cluster 2, on the other hand, represents stable low EPDOs. Despite the stability in EPDOs, the errors (MSE and RMSE) are higher than in Cluster 1 but lower than in Cluster 0 for ConvLSTM. This might be because while the data’s stability makes it easier to predict, the absolute errors remain low but perceptible, reflecting a consistent underestimation or overestimation by the models.

All region combined dataset was compared across models. When aggregating all clusters, it is notable that ConvLSTM and spatiotemporal ensembled-ConvLSTM perform significantly better than simpler models. The ensemble method likely leverages individual model strengths and mitigates their weaknesses, leading to improved overall prediction accuracy. The increasing complexity and adaptability of the models (from LR to ensembled-ConvLSTM) generally lead to better performance. ConvLSTM, integrating both convolutional and LSTM layers, efficiently handles spatial-temporal data, crucial for predicting EPDOs which are influenced by both spatial factors (e.g., road conditions and traffic density) and temporal factors (e.g., seasonal variations and time of day). The volatile nature of Cluster 1 might assist in model training by providing diverse scenarios for the models to learn from, which might explain the unexpectedly lower errors in this cluster compared to the more stable Cluster 2. In contrast, the stability in Cluster 2, while theoretically easier to predict, may lead to complacency in error reduction, resulting in slightly higher error metrics than Cluster 1. The superior performance of the spatiotemporal ensembled-ConvLSTM in ‘All regions’ suggests that combining multiple models helps capture a broader range of patterns and anomalies in the data, thereby enhancing prediction accuracy. This is particularly beneficial when dealing with heterogeneous data across various regions.

\section{Conclusion}\label{sec1}

Several conclusions and recommendations can be made regarding the use of the spatially ensembled-ConvLSTM framework for predicting weather-related crash risks. The spatially ensembled-ConvLSTM framework does outperform conventional predictive models, such as LR, ARIMA, and the standard ConvLSTM model in terms of accuracy. This is evidenced by the lower MSE and RMSE values across all regions, particularly when data from different crash risk zones are aggregated. The ensemble approach effectively combines the strengths of multiple ConvLSTM models, improving prediction accuracy through a robust handling of spatial and temporal variations in the data.

The proposed model exhibits distinct performance variations across different crash risk zones. In areas of volatile high risk (Cluster 1), the model achieves the lowest MSE and RMSE, suggesting a strong capability to handle and accurately predict scenarios with high variability in crash risks. Conversely, in stable low-risk areas (Cluster 2), the model still improves upon simpler models but shows slightly higher errors than in high-risk areas, likely due to the challenges in capturing subtle variations in inherently low-risk environments.

The framework’s ability to spatially align predictions with actual crash locations, especially noted in the superior performance in high-risk, volatile areas, indicates its logical validity. The integration of spatial data within the model allows it to effectively map and predict crash occurrences in relation to varying geographical and environmental factors, thereby confirming its utility and accuracy in practical applications.

While performance in high-risk areas is commendable, additional research and model tuning are recommended for low-risk zones to refine predictions and minimize errors. This could involve integrating more granular data points or exploring different ensemble configurations that are specifically optimized for stability rather than volatility. Additionally, to maximize the practical benefits of the spatially ensembled-ConvLSTM framework, integrating this model with real-time traffic monitoring systems could provide dynamic, timely predictions that can be directly utilized for traffic management and crash prevention. Finally, development of decision-support tools that leverage the model’s outputs to provide actionable insights for urban planners and public safety officials could significantly enhance the impact of the predictive capabilities.

\section*{Acknowledgements}

This paper is partially prepared based on information collected for a research project funded by the United States Department of Transportation - Office of the Assistant Secretary for Research and Technology (USDOT/OST-R) University Transportation Centers Program (Grant \#69A3551747127). 

\section*{Data Availability Statement}

The datasets generated and analyzed during this study are available from the corresponding author upon reasonable request. Additionally, this study utilizes raw data and pseudocode from Ogungbire \& Pulugurtha (2025), \textit{Spatiotemporal Risk Mapping of Statewide Weather-related Traffic Crashes: A Machine Learning Approach}, published in \textit{Machine Learning with Applications} (DOI: \href{https://doi.org/10.1016/j.mlwa.2025.100642}{10.1016/j.mlwa.2025.100642}). Access to these materials is subject to the policies of the original authors and the publisher.


\bibliography{sn-bibliography}

\end{document}